\documentclass[12pt]{article}
\usepackage{amsfonts,amsmath,amssymb,epsf}
\usepackage{graphicx,color}
\def\prt{\partial}

\def\d#1{\,{\rm d}#1}
\newcommand{\al}{\alpha'}
\newcommand{\de}{\partial}
\newcommand{\be}{\begin{equation}}
\newcommand{\ba}{\begin{eqnarray}}
\newcommand{\ea}{\end{eqnarray}}
\newcommand{\ee}{\end{equation}}

\newcommand{\we}{\wedge}
\newcommand{\ca}{\mathcal}

\newcommand{\f}{\frac}
\newcommand{\s}{\sqrt}
\newcommand{\vp}{\varphi}

\newcommand{\ap}{\alpha}

\newcommand{\ddd}{\cdot\cdot\cdot}
\newcommand{\no}{\nonumber \\}
\newcommand{\la}{\langle}
\newcommand{\lb}{\rangle}
\newcommand{\ep}{\epsilon}

\begin{document}
\begin{titlepage}
\thispagestyle{empty}
\begin{flushright}
UT-02-50 \\
HUTP-02/A045\\
hep-th/0209160 \\
September, 2002
\end{flushright}

\bigskip

\begin{center}
\noindent{\Large \textbf{
Notes on Giant Gravitons on PP-waves}}\\
\vspace{2cm}
\noindent{
Hiromitsu Takayanagi$^a$\footnote{hiro@hep-th.phys.s.u-tokyo.ac.jp} }
and \,
Tadashi Takayanagi$^b$\footnote{takayana@wigner.harvard.edu}
\\
\vskip 2.5em

{\it $^a$ Department of Physics, Faculty of Science,
University of Tokyo\\
Hongo 7-3-1, Bunkyo-ku, Tokyo, 113-0033, Japan\\
\noindent{\smallskip}\\
$^b$ Jefferson Physical Laboratory\\
Harvard University\\
Cambridge, MA 01238, USA\\}

\vskip 2em

\end{center}

\begin{abstract}
We investigate the giant gravitons in the maximally supersymmetric IIB
pp-wave from several viewpoints: (i) the dynamics of D3-branes,
(ii) the world-sheet description and (iii) the
correlation functions in the dual
N=4 Yang-Mills theory. In particular, we derive the BPS equation of a
D3-brane with magnetic flux, which is equivalent to multiple D-strings, and
discuss the behavior of solutions in the
presence of RR-flux. We find solutions which represent the excitations
of the giant gravitons in that system.

\end{abstract}
\end{titlepage}

\newpage

\section{Introduction}
\hspace{5mm}

Recently, it was shown that the string theory
on the maximally supersymmetric
type IIB pp-wave \cite{Bl1} can be obtained by taking
the Penrose limit of the familiar $AdS_5\times S^5$ \cite{Bl2}.
Furthermore,
the string theory on the pp-wave turns out to be exactly
solvable \cite{Metsaev,MeTs}.
Thus it is possible to check the AdS/CFT duality including the 
stringy modes and indeed this correspondence between the
gauge theory operators and perturbative closed string states has 
been successfully
compared \cite{BeMaNa}. However, there are also non-perturbative
objects
of D-branes which survive this limit and thus should also have
the gauge duals.
One of the important such
examples
is the giant graviton, which is described by a D3-brane
wrapped on $S^3$
\cite{GrSuTo,GrMyTa,HaHiIt}. The dual operator in the
N=4 Super Yang-Mills theory
was proposed in \cite{BaBrNaSt} and we can compute the correlation
function exactly \cite{CoJeRa}. As we will see later,
the correlation
function is finite in the large $N$ limit only if we
assume the scaling
of Penrose limit. Therefore we believe that the study
of giant gravitons
is important to know the duality between the pp-wave and
the gauge theory
well.

D-branes in pp-wave backgrounds can be classified into two groups
(for recent studies see [11]-[31]).
%\cite{BiPe,Dab,BeGaMaNaNa,Chu,LePa,KuNaSa,SkTa,BaHuLeNa,
%TaTa,Si,BaMeZa,AlKu,BeGa,Se,Me,Pa,Mi,AlGaGhPa,SuYo1,BePeZa,BiKuPa}).
One of them
can be described by the light-cone world-sheet theory 
\cite{BiPe,Dab,SkTa,BeGa}
and the other cannot \cite{SkTa,Se,Me}.
The Penrose limit of giant gravitons (spherical D3-brane)
is also divided into the two types above.
The open string spectrum
of the former D-branes
can be exactly examined and its comparison to the gauge theory
has been studied \cite{BeGaMaNaNa,LePa,SkTa,BaHuLeNa}. 
In the light-cone gauge
naively one may think that the D-brane cannot move away from
the origin because
of the mass term. However, as we will show later, some of the 
D-branes can
move without breaking supersymmetry
if we assume the appropriate boosting or gauge flux (for related
previous arguments see also \cite{SkTa,Mi}). In the case of a D3-brane
this just corresponds to the shift of the radius of giant gravitons.

On the other hand,
even though the latter
has no such stringy description, the D-brane in this group
has the interesting aspect that they can couple to the background RR-field
directly. For example, the D3-brane can be expanded into its spherical form
due to the RR-flux, which is identified with the giant graviton.
Another interesting example is the system of multiple D-strings. This can
couple to the RR-flux via the non-abelian term (Myers term \cite{My}).
As we will see in this paper, they can expand without losing any energy.
We will also investigate the system from the viewpoint of a D3-brane and
show that it is a 1/4 BPS state. These moduli of multiple 
D-strings are argued to
be excitations of giant gravitons.

This paper is organized as follows.
In Section 2 we discuss the world-volume of giant graviton 
in the Penrose limit
and also examine the correlators of its dual Yang-Mills operator.
In Section 3 we investigate the dynamics of multiple D1-branes 
and show that there
are BPS excitations of giant gravitons. In Section 4 we analyze 
the world-sheet
description of giant gravitons and derive the supersymmetric condition.
In the appendix we present the detailed computations of
the BPS condition of a D3-brane with magnetic gauge flux in the
pp-wave background.

\section{Giant Gravitons in PP-wave Background}\label{sec:giant}

The maximally supersymmetric pp-wave background\footnote{In
this paper we will use the notation of \cite{MeTs}
(but slightly different). In our
convention the SUGRA equation of motion is given by
$e^{-2\phi}R_{\mu\nu}=\f{1}{96}
F_{\ap\beta\gamma\delta\mu}F^{\ap\beta\gamma\delta}_\nu$.
This is the same as in Polchinski's text book \cite{Po} and has a merit
that WZ-term of
 D-branes is normalized conventionally
 such that $S_{WZ}=\mu_p\int_{Dp}C_{(p)}\ \
 (\mu_p=T_p e^\phi)$.} is given by
\ba
&&ds^2=-2\d x^+ \d x^- -\mu^2(\sum_{I=1}^8 x^I x^I )
(\d x^+)^2+\sum_{I=1}^8 (\d x^I \d x^I) , \no
&&\quad F_{+1234}=F_{+5678}=4\mu.
\ea
We can obtain this background from $AdS_5\times S^5$
\ba
ds^2=R_A^2\left(-dt^2\cosh^2 \rho+d\rho^2+\sinh ^2 \rho\ d\Omega^2_3
+d\psi^2\cos^2\theta+d\theta^2+\sin^2\theta d\Omega^{'2}_3\right),
\label{ads}
\ea
where the radius is given by $R_A=(4\pi g_sN{\al}^2)^{\f14}$, by
taking the Penrose limit $R_A\to\infty$ \cite{Bl2} scaling as
\ba
x^+=(t+\psi)/2\mu,\ x^-=R_A^2\mu (t-\psi),
\ \rho\to\rho /R_A,\ \theta\to\theta /R_A.
\ea
We also define the
polar coordinate of $S^5\ \ (Y_1^2+Y_2^2+Y_3^2+Y_4^2+Y_5^2+Y_6^2=R_A^2)$
as follows
\ba
&&Y_1=R_A\cos\theta \sin\psi,\  Y_2=R_A\cos\theta \cos\psi,\
Y_3=R_A\sin\theta\cos\ap \cos\beta,\no
&&Y_4=R_A\sin\theta\cos\ap \sin\beta,\
Y_5=R_A\sin\theta\sin\ap \cos\gamma,\ Y_6=R_A\sin\theta\sin\ap \sin\gamma.
\ea

\subsection{Giant Gravitons in the PP-Wave Background}

Here we would like to consider
giant gravitons (D3-branes) \cite{GrSuTo} in the Penrose limit.
A giant graviton is a D3-brane in the $AdS_5\times S^5$ background
which is rotated in the angular direction and is
expanded due to the force of RR-field. Its spatial
world-volume is given by $S^3$ in $S^5$ \cite{GrSuTo}
or $AdS_5$ \cite{GrMyTa,HaHiIt}. Since in the pp-wave background
we have the $Z_2$ symmetry which interchanges  $x^5\sim x^8$ with 
$x^1\sim x^4$ ( or equally in the AdS language $S^5$ with $AdS_5$),
the physical properties of these two
should be the same. Thus we only discuss
the first type of giant graviton here. Later we will mention that
it is not obvious, however, to see
this symmetry in the gauge dual picture.

Now, the resulting D3-branes in the Penrose limit can be largely
divided into two groups.
One of them is a spherical D3-brane
whose would-volume is taken in the
direction $x^+$ and $S^3(\in S^5)$.
After we take the limit,
this giant graviton is given by
\ba
t=\psi,\ \ a=R_A\theta(=\mbox{fixed}),
\ea
where $a$ is equal to the radius of $S^3$.

Another configuration
is the D3-branes which have Neumann boundary
conditions in the $x^\pm$,
$x^i$ and $x^j$ direction ($i,j\in\{1,2,3,4\}$ or $i,j\in\{5,6,7,8\}$).
In order to see this in detail, let us consider the giant
gravitons whose world-volume is given by
\ba
Y_1^2+Y_2^2+Y_3^2+Y_4^2={R'_A}^2,\ \  \gamma=t.
\ea
Next we take the Penrose limit $R_A\to\infty$
with keeping $r$ finite, where $r$ is defined
by
\ba
(R'_A/R_A)^2=1-(r/R_A)^2.
\ea

Then we obtain\footnote{Obviously we can shift the light-cone time
as $x^+\to x^++\mbox{const.}$ without changing physics.}
\ba
&&x^7=\s{y^2-r^2}\cos\beta, \ \ x^8=\s{y^2-r^2}\sin\beta,\no
&&x^5=r\cos(\mu x^+),\ \ x^6=r\sin(\mu x^+), \label{gai}
\ea
where we define the parameter $y=\theta R_A$. Since $\theta$ and $\beta$
are free parameters, the resulting D-brane has the Neumann boundary
condition
in the direction $x^7,x^8$ and the Dirichlet in $x^5,x^6$.
Thus this represents a D3-brane placed away from
the origin by $r$. Since the coordinates depend on $x^+$,
this brane is rotating in the $x^5-x^6$ plane.
If it does not rotate,
then it cannot preserve sixteen supersymmetries, as we will see later.

Below we first discuss the former type of giant gravitons.
The latter type allows the world-sheet analysis in the light-cone gauge
and will be studied in section 4.

\subsection{Comments on Correlators of Giant Gravitons in PP-wave Limit}

Here we would like to briefly comment on the gauge dual of the 
first type\footnote{The gauge dual of the
latter type, which allows the exact world-sheet analysis,
was discussed in \cite{BaHuLeNa} and the agreement of anomalous
dimension of their fluctuations is verified (see also \cite{DaJeMa}
for BPS fluctuations).} of
giant gravitons in the Penrose limit
employing the proposal of the dual Yang-Mills
operators discussed in \cite{BaBrNaSt}.
 
In the paper \cite{BaBrNaSt} the gauge theoretic dual of the
giant graviton with  angular momentum $J$ which expands in the
$S^5$ direction is conjectured to be
the BPS subdeterminant operator with $U(1)$ R-charge $J$
\ba
O^{S^5}_{J}=\f{1}{J!(N-J)!}\ep_{a_1a_2\ddd a_N}
\ep^{b_1b_2\ddd b_Ja_{J+1}\ddd a_N}Z^{b_1}_{a_1}\ddd Z^{b_J}_{a_J}.
\label{OS}
\ea
The radius $a$ of the corresponding giant gravitons is given by
\ba
a=\s{\f{J}{N}}R_A.
\ea
In particular the maximal one is described by the determinant operator.
Then by using the results and technique found
in \cite{CoJeRa},
we can compute
various (normalized) correlators of giant gravitons exactly\footnote{There
are several results on the
non-renormalization of extremal correlators e.g. \cite{nonren}
even for multitrace operators. Thus the results do not receive
no $g_{YM}$ corrections. We would like to thank S.Ramgoolam very much for
this argument. }. 
Now we are interested in the
Penrose limit $J\sim O(\s{N})$ and then the 
normalized three point function\footnote{The two point function, which 
is used to normalize other correlation functions, is given by
$\la \bar{O}^{S^5}_{J} O^{S^5}_{J} 
\lb= \f{N!}{(N-J)!}\sim N^J e^{-J^2/2N}$.}
in the N=4 $U(N)$ gauge theory is
given by
\ba
&&\la O^{S^5}_{J}\bar{O}^{S^5}_{J_1}\bar{O}^{S^5}_{J_2}\lb
=\s{\f{(N-J_1)!(N-J_2)!}{(N-J)!N!}}\simeq e^{-J_1J_2/2N}. \label{ts}
\ea
This result\footnote{The exponential behaviour may suggest a sort of a
tunneling effect. We would like to thank S.Minwalla very much for suggesting
this point and also other several useful comments.}
 shows that the single particle state of (\ref{OS}) is
well defined (i.e. not mixed with multiparticle states) if
we assume\footnote{More precisely, here we mean that $J^2/N$ is
always finite
taking the limit $N\to \infty$ and that the value $J^2/N$ is enough
larger than one.} $J^2/N \gg 1$.
Thus the condition that
the description of BPS operator (\ref{OS}) should describe the
geometrical object of giant graviton requires at least $J^2/N \gg 1$.
It may also be useful to note that the similar
behaviour of the correlator was found
in the pp-wave wave limit of $AdS_3\times S^3$ \cite{St}
(see also \cite{HiSu}).

Next let us compute the overlap between a single trace operator
and (\ref{OS}) in the limit $J\sim O(\s{N})$
\ba
&&\la \mbox{Tr}Z^J\ \bar{O}^{S^5}_{J}\lb\simeq
\f{e^{-\f{J^2}{4N}}}{\s{J}}
\s{\f{J^2/2N}{\sinh(J^2/2N)}}. \label{ova}
\ea
To derive the above result\footnote{The factor of $\sinh(J^2/2N)$ comes
from the the two point function \cite{KrPlSeSt,CoFrHeMiMoPoSk} 
$\la \mbox{Tr}Z^J \mbox{Tr}\bar{Z}^J \lb=JN^J\f{\sinh(J^2/2N)}{J^2/2N}$
, which is used to normalize.}
 we have 
used the fact that the operator $\mbox{Tr}Z^J$
can be represented  as a sum of Schur functions $\chi_h(Z)$, 
which is exactly the same
as the operator (\ref{OS}), such that $\mbox{Tr}Z^J=\sum_{h}\pm\chi_h(Z)$
(here $h$ denotes all of the hooks of Young diagram and 
for more detail see the appendix of the first paper of \cite{KrPlSeSt}). 

Let us try to interpret this result qualitatively
from the viewpoint of the
giant graviton in the string theory side. The single trace operator
is regarded as a light-cone vacuum $|p^+\lb$ following the
arguments in
\cite{BeMaNa}
and the wave function
in the supergravity is given by $\vp\sim e^{-\mu p^+x^2}$.
We also assume that the giant graviton has the width of $\delta a$ by
quantum effects such
that the wave function of D3-branes is
estimated by $\vp'\sim e^{-\f{(x-a)^2}{(\delta a)^2}}$.
Then it is easy
to see that the overlap (\ref{ova}) will be proportional to
$\exp(-\f{J^2/N}{\f{J(\delta a)^2}{R_A^2}+1})$. Thus we can find the
scaling
\ba
\delta a\sim \f{R_A}{\s{J}}=(4\pi{\al}^2 g_sN/J^2)^{\f14}\
(\ll a), \label{wi}
\ea
where we roughly examined neglecting
any finite constant factor only in this part. The width
becomes zero
if we neglect quantum effect $g_s=0$ as is expected. Also note that the
expression depends only on the effective coupling $\f{Ng_s}{J^2}$.
By using this result we can
reproduce the exponential
scaling behaviour of the
three point function (\ref{ts}).
Then we can argue that the description of the giant graviton by using the
operator (\ref{OS}) is
good\footnote{
The latter bound represents the critical point
 where gravitons become giant as
discussed in \cite{BeMaNa}.} if
\ba
(\delta a)^2 \ll \al \ll a^2.
\ea

There are also arguments on a giant graviton which expands in the
$AdS_5$ direction \cite{HaHiIt,CoJeRa}. The authors of \cite{CoJeRa}
conjectured that it corresponds to the operator
\ba
O^{AdS_5}_J=\f{1}{J!}\sum_{\sigma\in {\ca S}_{J}}Z^{a_1}_{a_{\sigma(1)}}
Z^{a_2}_{a_{\sigma(2)}}\ddd Z^{a_J}_{a_{\sigma(J)}},\label{OA}
\ea
where ${\ca S}_{J}$ denotes the permutation group of length $J$.

We can take the pp-wave limit for the normalized
three point function obtained in \cite{CoJeRa} as follows
\begin{equation}
\la O^{AdS_5}_{J}\bar{O}^{AdS_5}_{J_1}\bar{O}^{AdS_5}_{J_2}\lb
=\s{\f{(N+J-1)!(N-1)!}{(N+J_1-1)!(N+J_2-1)!}}
\simeq e^{J_1J_2/2N}. \label{ta}
\end{equation}

The overlap between a single trace operator and $O^{AdS_5}_{J}$ can also 
be computed in the same way as in (\ref{ova})
\ba
\la \mbox{Tr}Z^J\bar{O}^{AdS_5}_{J}\lb\simeq
\f{e^{\f{J^2}{4N}}}{\s{J}}
\s{\f{J^2/2N}{\sinh(J^2/2N)}}. \label{adsc}
\ea

As can be seen from these, both become too large when $J^2/N \gg 1$.
Thus
the expression (\ref{OA}) seems not be a good quantum state in 
our scaling limit. There is one more strong evidence for this.
In the pp-wave background there is the
${\bf Z_2}$ symmetry which exchanges $x^1\sim x^4$
with $x^5\sim x^8$ and
we should observe this symmetry in our correlation
functions. However, our
results do not show such a behaviour.

In the above we have examined several 
correlators of the gauge dual operators of
giant gravitons and found the exponential behaviour in the
pp-wave limit with
respect to the ones expanded in the $S^5$ direction.
Even though we gave
a quantitative interpretation of this result, the complete
explanation
requires the computation of the interaction between the
curved D-branes,
which is not tractable at present and will be left as a
future problem. The analysis of dual operators of
giant gravitons in $AdS_5$
seems to be much more difficult.

\section{Giant Gravitons in Multiple D1-branes on PP-Wave}

Next we would like to investigate giant gravitons in the pp-wave
from the viewpoint of lower dimensional D-branes (D1-branes).
Let us consider the D1-brane extended in 
the direction $x^+(\equiv \tau)$ and $x^4(\equiv y)$. 
The metric of its world-volume is given by
\ba
ds^2=-\mu^2y^2(dx^+)^2+dy^2.
\ea
This kind of D1-branes were first discussed in \cite{SkTa} and
shown to be 1/4 BPS. 
It is easy to check that it satisfies the equation 
of motion for DBI action in the pp-wave background. Below we are
interested in the static solutions of the non-abelian system of many 
D1-branes\footnote{For the relevant discussion of multiple D-string
in flat space see \cite{Di,Co}.
See also \cite{BrJaLoD} for discussions on
related non-abelian dynamics of D-strings
in more
general or other examples.}
(the number of D1-branes is defined to be $M (\gg 1)$).

\subsection{Non-abelian Dynamics of Multiple D1-branes on PP-Wave}

The transverse scalars are defined by
$\Phi^i=(2\pi\al)^{-1} x^i\ \ (i=1,2,3,5,6,7,8)$.
Below we only consider $\vec{\Phi}=(\Phi^1,\Phi^2,\Phi^3)$ because
the other scalars will not play any interesting
role in the presence of
the RR-field
$C_{+124}=-4\mu x^3$ via the RR couplings \cite{My}
$\mu_p\int \exp{(i2\pi\al\iota_{\Phi}\iota_{\Phi})}\we C
\sim\int \mbox{Tr}[C_{+ijk}[\Phi^i,\Phi^j]\de_{y} \Phi^{k}]$.
Then the non-abelian action which describes our static 
configuration is given by
\ba
S&=&-T_{D1}\int d\tau dy\mbox{STr}[\s{\mu^2(y^2
+(2\pi\al)^2|\vec{\Phi}|^2)(1+(2\pi\al)^2\de_y\Phi^i Q^{-1}_{ij}
\de_y\Phi^j)(\det Q_{ij})}\no
&&\pm T_{D1}\int d\tau dy \mbox{Tr}[(16i {\pi}^2 {\al}^2 \mu)
[\Phi^1,\Phi^2]\Phi^3],\label{ac1}
\ea
where we defined $Q_{ij}=\delta_{ij}+i(2\pi\al)[\Phi^i,\Phi^j]$ and 
the D-brane 
tension $T_{Dp}=\f{1}{(2\pi)^{p}g_s (\ap)^{(p+1)/2}}$.
The
symbol $\mbox{STr}$ denotes the symmetric trace \cite{Ts}. 
The last term
represents the effect of RR-flux \cite{My} and the $\pm$ sign
in front of it depends whether we consider branes or antibranes. We can 
choose the minus sign without losing the generality.
{}From this action
it is obvious that the constant shift of the
transverse scalars is not
the moduli of the system and it lifts the energy.
We are interested in
the way how to get the supersymmetric solutions with non-zero 
expectation values of the transverse scalars as shown below.

Since the full non-abelian action is too complicated, 
first we would like to assume
\ba
\al |\vec{\Phi}| \ll y, \ \ \ 
\s{\al} |\vec{\Phi}|\ll \s{M},\ \ \ \al|\de_y \vec{\Phi}|\ll 1,
\label{app}
\ea 
which restricts the value of $\vec{\Phi}$ at $y=0$ to zero.
Later we will
return to more general solutions.

Then we can approximate the action (\ref{ac1}) as follows
\ba
S&\cong& (\pi\al)^2 T_{D1}e^{-\phi}\int d\tau dy \mbox{Tr}
\Biggl[-2\mu |y||\de_{y}\vec{\Phi}|^2-\f{2\mu}{|y|}
|\vec{\Phi}|^2-16i\mu[\Phi^1,\Phi^2]\Phi^3\no
&+&\sum_{i,j}\mu |y|[\Phi^i,\Phi^j]^2\Biggr],\label{apps}
\ea
where we omit the constant term $ -\mu T_{D1}\int dy |y|$.
The validity of this approximation (\ref{app}) will be confirmed later.
We obtain the Hamiltonian as follows
\ba
H=\mu\pi^2{\al}^2 T_{D1}\int dy \ \ \mbox{Tr} 
\Bigl[2|y||\de_{y}\vec{\Phi}|^2+\f{2}{|y|}|\vec{\Phi}|^2
+\f{8}{3}i\ep_{ijk}[\Phi^i,\Phi^j]\Phi^k-|y|{[\Phi^i,\Phi^j]}^2\Bigr].
\label{ac}
\ea

We can rewrite the expression (\ref{ac}) as a sum of
squares and boundary
terms
\ba
H&=&\mu\pi^2{\al}^2 T_{D1}\int dy \ \ \mbox{Tr} 
\Biggl[\f{2}{|y|}\left(y\f{d\Phi^i}{dy}-\Phi^i-
\f{i|y|}{2}\ep_{ijk}[\Phi^j,\Phi^k] \right)^2\no
&+&\f{\de}{\de y}
\left(2\f{y}{|y|}\Phi^i\Phi^i
+\f{2iy}{3}\ep_{ijk}[\Phi^i,\Phi^j]\Phi^k\right)\Biggr],\label{BPS}
\ea
where we have used the previous constraint $\Phi^i(0)=0$.
In this way we obtain the following BPS equation (or\footnote{See 
also \cite{BaHoPi} 
for a different appearance of Nahm like equation from the 
D1-branes \cite{Di,Co} in AdS background, which is not related to our
D1-branes directly.}
`Nahm like equation 
in the pp-wave 
background')
\ba
\f{d \Phi^i}{dy}=
\f{i}{2}\f{y}{|y|}\ep_{ijk}[\Phi^j,\Phi^k]+\f{1}{y}\Phi^i.\label{Nahm}
\ea

To find its solution we assume the following ansatz 
\ba
[\Phi^i,\Phi^j]=i\ep_{ijk}R(y)\Phi^k,\label{r}
\ea
that is the fuzzy sphere structure \cite{fuzzy}.
Since we are interested in a bound state
of $M$ D1-branes, we have only to consider the $M$ dimensional 
irreducible representation of the $SU(2)$ algebra. In the most of the
discussion below we assume $M$ is large.
Then its trace is normalized such that
\ba
\sum_{i=1}^{3}\mbox{Tr}(\Phi^i\Phi^i)=\f{(M-1)M(M+1)}{4}R(y)^2,
\ea
and thus the radius $r$ of the sphere is given by
\ba
r=2\pi\al\s{\f{1}{M}\mbox{Tr}[\Phi^i\Phi^i]}
=\s{M^2-1}\pi\al R\sim M\pi\al R.
\ea

Now, the nontrivial solutions of (\ref{Nahm}) are obtained as follows
\ba
R(y)=\f{2C|y|}{1+Cy^2},\label{bs}
\ea
where $C$ is an arbitrary constant.
If $C$ is negative, then the value of $R$ diverges at some points and 
this solution may seem to represent a kind of a spike solution.
However, in this case the
approximation (\ref{app}) is violated and indeed later we will see that
there is no such BPS solution from the view point of a D3-brane with 
magnetic flux.

Therefore we regard $C$ as a positive constant. Then 
we find the smooth solution in $-\infty<y<\infty$.
Its energy is given by zero (see (\ref{BPS})) and thus it is 
expected to be stable and BPS. In other words the deformation of the
original D1-branes by the parameter $C$ should be 
the moduli of supersymmetric vacua. Though it is not easy to check the 
supersymmetry in this non-abelian system, we will show that this 
is the case in the analysis of the D3-brane.
Our approximation of (\ref{app}) is valid if
 $MC\al \ll 1$ for the BPS 
solution (\ref{bs}) (remember we are interested in the
limit of $M \gg 1$.)
Then the maximal radius of the fuzzy sphere 
can also be estimated $r\sim \pi\al M\s{C}$ and can
take larger values than the
string scale.
Thus we can believe this solution is physically relevant.
We would also like
to note that all the computation above does not use the condition 
$M\gg 1$ and
even if we do not assume it, the solution 
(\ref{bs}) may be meaningful.

Next we discuss the interpretation of this classical solution.
Since 
the $M$ D1-branes can expand into a D3-brane due to Myers effect
\cite{My}, we can speculate that the configuration (\ref{bs})
is a BPS 
bound state of $M$ D1-branes and two D3-branes 
each located at $y=\pm 1/\s{C}$. These two D3-branes can be regarded as
a giant graviton with the small radius $r\ll \s{M\al}$
at the origin divided into two parts.
As we will see in the 
next subsection we can show that the $1/4$ BPS solutions of a D3-brane
with magnetic flux include this solution.
Our result that the energy of this system is the same as that of 
$M$ D1-branes is also consistent with the fact that giant
gravitons have
the vanishing value of $p^-=\mu(\Delta-J)$. 

\subsection{BPS Configurations of D3-brane in PP-wave Background}

Now one may ask what will happen if we consider
larger giant graviton $r\gg \s{M\al}$ in the pp-wave background.
In order to examine them we must return to the 
full non-abelian action (\ref{ac1}). Since we assume
$M \gg 1$, the symmetric
trace STr can be replaced with the ordinary trace Tr. 
Applying the ansatz (\ref{r}), we can rewrite the Hamiltonian of the 
action (\ref{ac1}) as
follows ($\ap^2\equiv(2\pi\al)^2(M^2-1)$)
\ba
H=M\mu T_{D1}\int dy[\s{(y^2+\f{\ap^2}{4}R^2)(1+\f{\ap^2}{4}R^4)
(1+\f{\ap^2}{4}(\de_y R)^2)}-\f{\ap^2}{3}R^3].\label{R1}
\ea
This action just coincides with the effective action
(DBI+Chern-Simons term)
of a D3-brane with the magnetic flux on $S^2$ (fuzzy sphere)
\ba
F=(\pi\al M)\sin\theta d\theta\we d\vp,
\ea
in the pp-wave background. Then we can derive the supersymmetric 
condition (BPS equation) of this system since we have only to consider
a single D-brane. As we show the detailed analysis in the appendix,
we can verify that the system keeps eight
dynamical supersymmetries (1/4 BPS), which is the same number in D1-branes,
 if the following BPS equation is satisfied
(see (\ref{BPSEQ}) and use the relation $\rho=\ap R/2$) 
\ba
\f{dR}{dy}=\f{\pm R^2 y+R}{y\mp \f{\ap^2}{4}R^3},
\label{BPSD3}
\ea
where we should choose $+\ (-)$ sign in the right-hand side
if $y+ \f{\ap^2}{4}R\, \prt_y R<0$ ($>0$).
We can rewrite this as $y-\frac{\ap^2}{4}R^3<0$
$(y+\frac{\ap^2}{4}R^3>0)$ by using the BPS equation.
The existence
of dynamical supersymmetries $Q^-$ 
is very important because we now regard
the energy as the light-cone energy $p^-(\sim\{Q^-,Q^-\})$.
Note also that if we assume the approximation $|y|\gg \f{\ap^2}{4}R^3$
(or equally
$MC\al\ll 1$ as before), we get the previous BPS equation
\ba
\f{d R}{dy}=\pm R^2+\f{R}{y}.
\ea
This just matches with our previous result of non-abelian D1-branes.
If we consider the D3-brane with no magnetic flux $M=0$, then we can 
also see
that the solution is given by a three dimensional sphere
$\rho^2+y^2=$constant and this is consistent with the result
in \cite{BeMaNa,SkTa}. 

This BPS equation can also be obtained from the
viewpoint of the action of a D3-brane (or equally multi D-strings) (\ref{R1}).
For this aim the crucial identity is
\ba
&&(y^2+\f{\ap^2}{4}R^2)(1+\f{\ap^2}{4}R^4)
(1+\f{\ap^2}{4}(\de_y R)^2)\no
&&=\left(\f{\ap^2}{4}(yR^2R'-R^3)\pm\! (y+\!\f{\ap^2}{4}RR')\right)^2\!
+\!\f{\ap^2}{4}\left((yR^2+\!\f{\ap^2}{4}R^3R')\mp (yR'-R)\right)^2\!.
\ea
Note that the second term is the square of BPS eq.(\ref{BPSD3}).
Then the Hamiltonian (\ref{R1}) is estimated as
follows (we restore $\mu$)
\ba
H\geq \f{\ap^2M}{4}T_{D1}
\mu\int dy\f{d}{\de y}\left(-\f{1}{3}R^3 y\mp
\f{1}{2}
R^2\right)\mp MT_{D1}\mu \int dy\ y.
\ea
The equality holds only when BPS eq.(\ref{BPSD3})is satisfied.
The second term represents the energy of $M$ D1-branes
with $\pm$ signs. If we assume that $R(0)=0$
, $lim_{y\to\infty}y R(y)^3=0$ and $lim_{y\to\infty}R(y)^2=0$ (the
second and third condition mean that we consider
a solution which is asymptotically D1-branes),
then we obtain\footnote{This is because the condition
$y-\f{\ap^2}{4}R^3<0$ ($y+\f{\ap^2}{4} R^3>0$) is equivalent
to $y<0$ ($y>0$) if $R(0)=0$.}
\ba
&&H-MT_{D1}\mu\int dy |y|\no
&&\geq \f{\ap^2M}{4}T_{D1}
\mu\left[\int^{\infty}_{0} dy\f{d}{\de y}\left(-\f{1}{3}R^3 y+
\f{1}{2}R^2\right)+
\int^{0}_{-\infty}dy\f{d}{\de y}\left(-\f{1}{3}R^3 y-
\f{1}{2}R^2\right)\right]
\no
&&=0.\label{zeroe}
\ea
Thus in this case all of the BPS solutions have the same energy as
$M$ D1-branes. In other words this is the moduli of the system.

Now we would like to classify the solutions to the 
BPS equation (\ref{BPSD3}).
After some numerical study we find that there are
two types of solutions.
The first type is well approximated by the previous
solutions (\ref{bs}). This exists only when $R_{max}<R_0$,
where $R_0$
is a constant of order $\sim \ap^{-\f12}$. See Fig.\ref{Fig1} below.
All of these solutions have the zero energy as we can see from
 (\ref{zeroe}). Note that $R(0)=0$ holds
for them and they are symmetric under $y\to -y$.

\begin{figure}[htbp]
  \centerline{\epsfbox{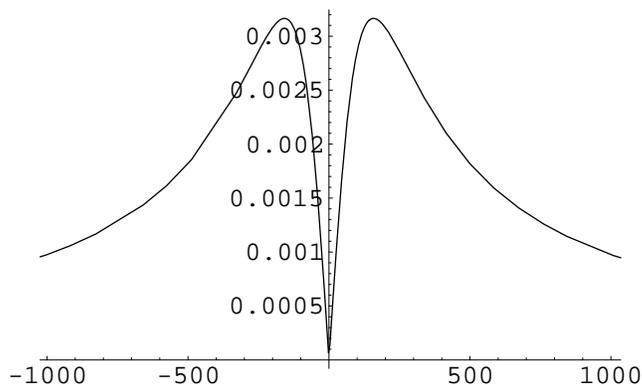}}
	\caption{The behaviour of a solution $R(y)$ in the case 
	$R_{max}(=0.032)<R_{0}$. Here we scaled $y$ and $R(y)$ 
	such that $\ap=1$.}
	\label{Fig1}
\end{figure}

Another type of solutions is possible for $R_{max}>R_0$ and this is not exactly
symmetric under $y\to -y$. This includes a large lump at
the origin, which is
regarded as a giant graviton. See Fig.\ref{Fig2}
below. The remarkable result is that this
solution
has less energy than the original D1-branes. 
Thus this implies that the system of
multiple D1-branes
may be unstable, while a single D1-brane is stable.
The `instability' observed here is closely related to the fact that
even though
the eight Killing spinors always are preserved
in the D3-brane system, they are not continuous at the
point $y+\f{\ap^2}{4} R^3=0$ as the $\pm$ sign in (\ref{1234}) flips.
In this sense the 1/4 BPS system of the D3-brane
in the pp-wave background is somewhat different from the usual
BPS state. This nature of Killing spinors
also equally applies to a single D1-brane
in the pp-wave background in the same way even though this is not mentioned
before. In a sense we may say that
the D3-brane (or D1-brane)
turns into anti D3-brane (anti D1-brane) at the point 
$y+\f{\ap^2}{4} R^3=0$.
However, we have some doubts whether this solution is physical or not
because for this solution
we have the unnatural discontinuity\footnote{
Even though the first solution (Fig.\ref{Fig1}) has also a
discontinuity at $y=0$, the D3-brane world-volume can be described by the 
`double cone' and is not unnatural. It is also easy to see that the
discontinuity does not contribute to the energy by an 
explicit calculation.} of 
the differential of $R$ at the point $y+\f{\ap^2}{4} R^3=0$. We would 
like to leave the correct 
interpretation of this solution as a future problem. If this instability is 
true, then our result tells that a system of multiple D1-branes will
decay to a giant graviton, which has the zero energy.

\begin{figure}[htbp]
  \centerline{\epsfbox{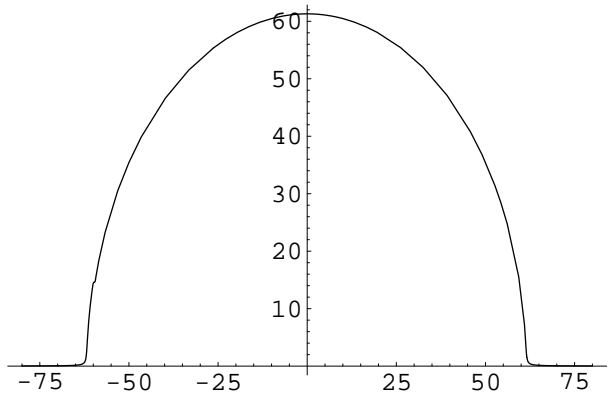}}
	\caption{The behaviour of a solution $R(y)$ in the case 
	$R_{max}(=62)>R_{0}$. Here we scaled $y$ and $R(y)$ such that $\ap=1$
	as before. This solution is not exactly left-right symmetric in spite of 
	its appearance. The derivative of $R(y)$ is not continuous at 
	$y=y_0 ( \simeq -59.5)$.
	The region $y>y_0$ corresponds to the minus sign in (\ref{BPSD3})
	and the one $y<y_0$ to the plus sign.}
	\label{Fig2}
\end{figure}

\section{World-sheet Description of Giant Gravitons}
\setcounter{equation}{0}
\hspace{5mm}
In this section we investigate those BPS D-branes on the pp-waves
which can be expressed in the light-cone string formalism. Especially,
this includes
the Penrose limit of giant gravitons and we will examine them in detail.

In the type IIB theory on the flat background, the BPS D-branes 
are (D(-1),D1,D3,D5,D7,D9) and we can put
these D-branes at any points preserving 16 supersymmetries out of 32. 
On the other hand, in the pp-wave background,
which has also 32 supersymmetries, all of these D-branes
are not BPS due to its curved
background and RR-flux \cite{Dab} (see also
\cite{BiPe,BeGa,BePeZa}).
Naively, BPS D-branes are pinned at the origin
because the RR-flux acts as the mass term. However, it can move around without
losing the supersymmetries if we assume gauge flux on D-branes
as first pointed out in the example of D5-brane
\cite{SkTa} (see also \cite{Mi} for the similar issue on
BPS D-branes in the PP-wave limit of
$AdS_3\times S^3 \times T^4$). This is very natural since this background is
Lorentzian symmetric space in spite of its appearance in the light-cone gauge. 
Below we conduct the complete investigation of supersymmetric D-branes
on the pp-waves
in terms of the open string
boundary conditions, which includes a series of the
Penrose limit of giant gravitons as D3-branes.

The string action on this pp-wave background can be described
in the GS formalism in the light cone gauge \cite{Metsaev}.
Then we are restricted to open strings between D-branes
whose boundary conditions in the
light cone directions $X^+,X^-$ are both Neumann, i.e.,
$(+,-,m,n)$ D-branes. The notation $(+,-,m,n)$ means the D-brane
whose Neumann Directions are the two light-cone directions $X^+,X^-$,
$m$ ones in $X^1,\cdots,X^4$
and $n$ ones in $X^5,\cdots,X^8$ following the convention in \cite{SkTa}.
In the light-cone
gauge the original 32 supersymmetries are divided into 16 kinematical and 16
dynamical supersymmetries, where the half BPS D-branes in the
light-cone gauge preserve 8 out of each.
Under this condition we
can find that D-branes
which preserve partial supersymmetries are {\it only} those in the table below
\cite{Dab,SkTa}\footnote{It was pointed out in \cite{BeGa} 
that the D-branes which preserve less than half supersymmetries
are inconsistent because they do not seem to have the open-closed modular 
property.}.

%\begin{table}[h]
%\vspace{-12pt}
\begin{center}
\begin{tabular}{|c|c|c|c|c|}
\hline
&World-Volume&Dynamical SUSY&Kinematical SUSY&Total\\
\hline
D1 & $(+,-,0,0)$&8&0&8\\ \hline
D3 &$(+,-,2,0),(+,-,0,2)$&8&8&16\\ \hline
D5 &$(+,-,3,1),(+,-,1,3)$&8&8&16\\ \hline
D7 &$(+,-,4,2),(+,-,2,4)$&8&8&16\\ \hline
\end{tabular}
\end{center}
%\caption{}\label{table BPS brane}
%\end{table}
Except D1-branes, all dynamical supersymmetries are broken if the
D-branes are away from the origin. On the other hand, the D1-branes
keep $8$ dynamical ones even if these are away from the origin.

However, it
may be possible that these D-branes can be moved away from the origin
with preserving their 8 dynamical SUSY by adding the gauge flux on them
or their boosting. In our approach, the flux and boost can be treated easily
because these appear only as the change of the boundary condition. We will
find in section \ref{BPS D-branes sub} that the result is described by
the table below \footnote{Though we treat only the 
D-branes of the special directions in this table
for simplicity, the general results can be easily reproduced 
by the background $SO(4)\times SO(4)$ and $Z_2
\, (1234 \leftrightarrow 5678)$ symmetry.}.
%\begin{table}[h]
\vspace{-12pt}
\begin{center}
\begin{tabular}{|c|c|c|c|c|c|}
\hline
&World-Volume&Dynamical SUSY&Total&Condition&Movable directions\\
\hline
D3 &$+,-,7,8$&8&16&boost in $5,6$&$5,6$\\ \hline
D5 &$+,-,4,6,7,8$&8&16&flux in $4$&5\\ \hline
D7 &$+,-,3,4,5,6,7,8$&4&12&boost or flux&$1,2$\\ \hline
\end{tabular}
\end{center}
%\caption{}\label{table BPS brane fv}
%\end{table}
The symbol `flux in $I$' means the addition of the proper gauge flux $F_{+I}$ 
on the D-brane.
A D5-brane with the gauge flux 
can be placed away from the origin keeping 16 supersymmetries 
as already pointed out 
in \cite{SkTa} in a somewhat different approach. Our new result here is 
that a half BPS D3-brane is also movable with the appropriate boosting.
We will find that this rotating D3-brane corresponds to the giant graviton
which is discussed in the section \ref{sec:giant}.
 On the other hand, a half BPS D7 is not movable without breaking 4
 supersymmetries (movable D7 totally preserves 12 ones).

We can also see the effect of the gauge flux or boosting 
in the open string spectrum.
When we consider the 1/2 BPS D-brane without flux and boost,
we can find that the open string spectrum on this brane
has the mass term $\sim \mu^2x^2$ when the brane is moved to $x\neq 0$.
On the other hand, when we consider 1/2 BPS rotating D3-brane, 
the open string spectrum which is calculated in the rotating
coordinates \cite{Mich} does not have the mass term. 
We can also obtain the similar
result in the case of the 1/2 BPS D5-brane with the proper flux.
It would be very interesting if we can compare this spectrum with the 
gauge theoretic results.

\subsection{String action and Supersymmetry}
We mainly follow the notation of Metsaev and Tseytlin \cite{MeTs}.
The string world-sheet action is 
\begin{equation}
\begin{split}
S&=\frac{1}{2\pi \alpha^{\prime}}\int \d \tau \int_{0}^{\pi}
\d \sigma \Big[ \frac{1}{2} \prt_+ X^I \prt_- X^I -\frac{1}{2}m^2 (X^I)^2\\
&\qquad +\sqrt{2}\big[i(\bar S \prt_{\tau} S + 
S \prt_{\tau} \bar S
+S \prt_{\sigma} S +\bar S 
\prt_{\sigma} \bar S)
-2m\bar S \Pi S\big] \Big],
\end{split}\label{action S}
\end{equation}
where we use the light cone gauge $X^+=p^+ \tau $
and define $m=\mu p^+$, $\prt_{\pm}=\prt_{\tau}\pm \prt_{\sigma}$.
$S$ is an 8 dimensional complex spinor and $\bar S$
is its conjugate. We defined $\Pi\equiv \gamma^1\gamma^2\gamma^3\gamma^4$,
where $\gamma^I$ is the real and symmetric $SO(8)$ $\gamma$ matrix.
The equations of motion are
\begin{equation}
\begin{split}
&\qquad \qquad \prt_+ \prt_- X^I + m^2 X^I =0,\\
&\prt_+ S^1 - m\Pi S^2=0,\quad \prt_- S^2 +m\Pi S^1=0,
\end{split}
\end{equation}
where we defined real spinors $S^1$ and $S^2$ as follows
\begin{equation}
S=\frac{1}{\sqrt{2}}(S^1+iS^2),\quad \bar S=\frac{1}{\sqrt{2}}(S^1-iS^2).
\end{equation}
Finally, the dynamical supersymmetry transformation is given by
\begin{equation}
\begin{split}
\delta_{\epsilon}X^I&=-2i\Big[\epsilon^1\gamma^IS^1
+\epsilon^2\gamma^IS^2\Big],\\
\delta_{\epsilon}S^1&=\frac{1}{\sqrt{2}}\Big[
\prt_-X^I \epsilon^1\gamma^I+mX^I\epsilon^2\gamma^I\Pi \Big],\\
\delta_{\epsilon}S^2&=\frac{1}{\sqrt{2}}\Big[
\prt_+X^I \epsilon^2\gamma^I-mX^I\epsilon^1\gamma^I\Pi \Big],
\end{split}\label{Dynamical S1}
\end{equation}
and the kinematical supersymmetry transformation is also 
\begin{equation}
\delta_{\kappa} S^1=\cos m\tau \kappa^1+\sin m\tau \Pi \kappa^2 ,\quad
\delta_{\kappa} S^2=\cos m\tau \kappa^1-\sin m\tau \Pi \kappa^2.
\label{kin susy trans}
\end{equation}

\subsection{BPS D-branes on the PP-wave Background}
\label{BPS D-branes sub}
Let us consider the open string whose both endpoints are attached to
the same D-brane. The boundary conditions are 
the Neumann condition (N) and the Dirichlet condition (D)
\begin{equation}
N:\quad X^{\prime I}=0,\quad 
D:\quad \dot X^I=0,\quad \mbox{at} \quad \sigma=0,\pi,
\label{BC boson}
\end{equation}
where  we defined $\dot X^I \equiv \prt_{\tau}X^I$ and
$X^{\prime I}\equiv \prt_{\sigma}X^I$.
Next, the boundary condition of fermionic fields is given by
\begin{equation}
S^1|_{\sigma=0,\pi}=\Omega S^2|_{\sigma=0,\pi},
\end{equation}
where $\Omega$ is constructed
by multiplying $\gamma^i$ in the Dirichlet directions as in the case of
the flat background. Then we can count the number of SUSY charges which
is preserved on the D-brane,
by counting the number of $\epsilon$ and $\kappa$ which are chosen
so that $\delta S^1$ and $\delta S^2$ generated by $\epsilon, \kappa$
satisfy the relation
\begin{equation}
\delta S^1|_{\sigma=0,\pi}=\Omega \delta S^2|_{\sigma=0,\pi}.
\label{SUSY cond}
\end{equation}\\

{\bf{Kinematical SUSY}}

First let us consider the kinematical SUSY.
By using eq.(\ref{SUSY cond}) and (\ref{kin susy trans}), we find
\begin{equation}
\begin{split}
&\quad \cos m\tau (\kappa^1-\Omega \kappa^2)+\sin m\tau 
(\Pi \kappa^2+\Omega \Pi \kappa^1)=0\\
&\Rightarrow \kappa^1=\Omega \kappa^2, \quad
(1 + \Pi\Omega \Pi \Omega)\kappa^2=0.
\end{split}\label{kappa1}
\end{equation}
Then, only the D-branes which satisfies
$\Omega \Pi \Omega \Pi=-1$ always preserve 8 kinematical SUSY and
the other D-branes break all kinematical SUSY.
As discussed in \cite{Dab}, the choices of 
$\Omega$ which satisfies $\Omega \Pi \Omega \Pi=-1$
are only 
\begin{equation}
\begin{split}
D7&: \quad \gamma^i\gamma^j, \quad \gamma^{i^{\prime}}\gamma^{j^{\prime}},\\
D5&: \quad \gamma^{i^{\prime}}\gamma^i\gamma^j\gamma^k,\quad
\gamma^i\gamma^{i^{\prime}}\gamma^{j^{\prime}}\gamma^{k^{\prime}},\\
D3&: \quad \gamma^{i^{\prime}}\gamma^{j^{\prime}}\Pi , \quad
\gamma^i\gamma^j\Pi^{\prime},
\end{split}\label{bps}
\end{equation}
where we defined $i,j,k=1,\cdots,4$, $i^{\prime},j^{\prime},k^{\prime}
=5, \cdots ,8$ and 
$\Pi^{\prime}\equiv \gamma^5\gamma^6\gamma^7\gamma^8$.
Notice that D1 ($\Omega=\Pi\Pi^{\prime}$) and D9
($\Omega=1$) break all kinematical SUSY.\\

{\bf Dynamical SUSY}

Next let us consider dynamical SUSY. By using
eq.(\ref{Dynamical S1}),(\ref{BC boson}) and (\ref{SUSY cond}),
we obtain the condition
\ba
(\prt_-X^I\gamma^I\epsilon^1+m X^I\Pi \gamma^I \epsilon^2)
=\Omega (\prt_+ X^I \gamma^I \epsilon^2 -m X^I \Pi \gamma^I \epsilon^1)
,\quad \mbox{at} \quad \sigma=0,\pi 
\ea
and this is rewritten as follows
\ba
\quad (\dot X^a\gamma^a-X^{\prime i}\gamma^i)(\epsilon^1-\Omega \epsilon^2)
+mX^I(\Pi\gamma^I\epsilon^2+\Omega\Pi\gamma^I\epsilon^1)=0,
\label{SUSY cond dyna}
\ea
where we have used the index `$a$' as Neumann directions
and `$i$' as Dirichlet. This relation must be satisfied at any
$\dot X^{a},X^{\prime i}$ and $X^a$ and thus we obtain
\ba
\epsilon^1=\Omega \epsilon^2,\quad
mX^a\Pi(1+\Pi\Omega\Pi\Omega)\gamma^a\epsilon^2
+mX^i\Pi(1-\Pi\Omega\Pi\Omega)\gamma^i\epsilon^2=0
%&\Rightarrow \quad
%\epsilon^1=\Omega \epsilon^2,\quad \Pi\Omega\Pi\Omega=-1,\quad X^i=0
\label{SUSY cond dyna2}
\ea
is the condition. Namely, only the
D-branes which satisfy $\Pi\Omega\Pi\Omega=-1$
and $X^i=0$ preserve half dynamical SUSY and otherwise
all dynamical SUSY is broken.

Notice that D1-brane ($\Omega=\Pi\Pi^{\prime}$)
is exceptional because there is no $X^a$
direction. D1-brane satisfies eq.(\ref{SUSY cond dyna2}) at
any $X^i$, i.e., half dynamical SUSY is preserved at any $X^i$.\\

{\bf Dynamical SUSY with Gauge flux and boost}

As we saw, D3,D5 and D7-branes which
satisfy $\Pi\Omega\Pi\Omega=-1$ preserve 8 dynamical SUSY charges only when
these are at the origin. It may be possible, however,
that these D-branes are moved from the origin
with preserving their 8 dynamical SUSY by adding the gauge flux and boosting, 
because we can change the boundary condition with the gauge flux and boost.
With the gauge flux $F_{\mu \nu}$ and boost $v_i$,
we can shift the boundary condition to
\begin{equation}
0=X^{\prime a}-F_{a+}\prt_{\tau}X^{+}=
X^{\prime a}-p^+F_{a+},\quad
0=\dot X^i -v_i\prt_{\tau}X^{+}=\dot X^i -p^+v_i,
\end{equation}
and then eq.(\ref{SUSY cond dyna2}) changes to
\begin{equation}
\epsilon^1=\Omega\epsilon^2,\quad \Big[A_I\gamma^I\Omega
+mX^i\Pi\gamma^i \Big]
\epsilon^2=0, \quad A_I\gamma^I \equiv p^+(-F_{a+}\gamma^a+v_i\gamma^i).
\label{SUSY cond dyna fv}
\end{equation}
All we have to do is to study this equation for each of
D3,D5,D7-branes. Let us study D3-brane first
because we are especially interested
in giant gravitons.

When D3-brane is extended in $+,-,7,8$ directions
($\Omega=\Pi\gamma^5\gamma^6$), eq.(\ref{SUSY cond dyna fv}) is
\begin{equation}
\epsilon^1=\Omega\epsilon^2,\quad \Big[A_I\gamma^I
+m\gamma^5\gamma^6(X^1\gamma^1+\cdots+X^6\gamma^6)
\Big]\epsilon^1=0.
\label{D3flux1}
\end{equation}
Eq.(\ref{D3flux1}) is satisfied for any $\epsilon^1$ when we set
\begin{equation}
A_5=-mX^6,\quad A_6=mX^5,\quad X^1=X^2=X^3=X^4=0.
\end{equation}
Namely, D3-brane can move $X^5,X^6$ directions freely
with preserving the half SUSY by boosting in the
$X^5$ and $X^6$ directions. We can interpret this D-brane 
as a D3-brane rotating in the $x^5-x^6$ plane. This D-brane just coincides
with the Penrose limit of the giant graviton (\ref{gai}) 
discussed\footnote{With respect to the D3-brane at the origin 
this interpretation was first discussed in \cite{BaHuLeNa}. 
Here we extend this
result into that away from the origin.}
 in 
the section \ref{sec:giant}. 
The ratio of the radius $r$ to velocity $v$ is given by 
\ba
r/v=\s{(X^5)^2+(X^6)^2}/v=\f{1}{\mu},
\ea
and this is indeed the same as in (\ref{gai}).

The issue of whether the brane is movable or not will also be important
when we consider orbifold theories of the pp-waves. If we consider the 
Penrose limit of 
$AdS_5\times S^5/Z_M$ orbifold as in \cite{orbpp}, then we can naturally
identify each of $M$ types of fractional D3-branes \cite{DiDoGo}
in the pp-wave with the dibaryon
operators 
\ba
B^i=\det Q_{i},\ \ (i=1,2,\ddd,M)
\ea
where $Q_{i}$ denotes the complex scalar in the direction
$x^5-x^6$ which corresponds to the 
$i$-th arrow of the quiver diagram. Note that the subdeterminant
operator like (\ref{OS}) is not allowed in this case because of the requirement
of the gauge invariance. This is consistent with the fact that fractional
D-branes are pinned at the fixed point. If we consider the superposition of
$M$ different fractional D-branes, then we can put it away from the 
fixed point. The gauge dual of this fact is that we can construct the 
(gauge invariant) 
determinant operator $\det Q=\prod_{i=1}^{M}\det Q_{i}$ and
the subdeterminant operators. It may also be interesting to 
consider the correspondence between open string in the pp-wave 
and gauge theory as in \cite{BeHeKl,BaHuLeNa}.

We can also examine D5 and D7-branes similarly. A D5-brane
($\Omega=\gamma^5\gamma^1\gamma^2\gamma^3$)
with the gauge flux $p^+F_{4+}=mX^5$ can move in the $X^5$ direction
preserving 8 dynamical SUSY.
This fact was already pointed out in terms of the world-volume theory
in \cite{SkTa} and can be obtained the Penrose limit of $AdS_4\times S^2$
brane found in \cite{KaRa}.
A D7-brane with any flux and boost
can not move from the origin with preserving 8 dynamical
SUSY, but it is easy to see that 
preserving 4 dynamical SUSY is possible.

\section{Conclusions and Discussions}
\hspace{5mm}
In this paper we have examined the giant gravitons in the pp-wave background
from several viewpoints. In the background they can be described by
spherical D3-branes whose world-volume includes only $x^+$ direction or
D3-brane planes whose world-volume includes both 
$x^+$ and $x^-$. With respect to
the former giant graviton we computed the
correlators in the dual gauge theory and found that they are
finite in this limit and show their interesting exponential behaviour.

We also see
that the former excitation of giant graviton
can appear in the system of multiple D1-branes and indeed
they give moduli of the non-abelian D1-branes. It would be interesting to
study the dual Yang-Mills picture of this phenomenon.

The latter giant graviton can be described by the world-sheet theory
in the light-cone gauge. We showed that the corresponding D3-brane should be
boosted in order to preserve sixteen supersymmetries. 
This result just matches
with the Penrose limit of giant gravitons. We also computed the open
string spectrum and observed that the zero-mode mass term vanishes.

There is another world-sheet description of
D-branes, i.e. boundary states \cite{BiPe,BeGa}. Since all of these 
D-branes are
space like in the light-cone gauge, they are completely different from those
discussed
in this paper. Because the boundary state can be expanded 
in terms of closed
string modes, we may regard it as a coherent sum of BMN operators. 
The intriguing point is that even though
the massive modes
are included in the boundary states, they can preserve 
the partial supersymmetry.
It would be interesting to understand the gauge theoretic
interpretation of this.

\vskip2mm
{\bf Acknowledgments}

\vskip2mm
We would like to thank N.Constable, M.Hamanaka, Y.Hikida, Y.Imamura,
T.Kawano, S.Minwalla, L.Motl,
A.Strominger and N.Toumbas for useful discussions and comments.
We are also grateful
to S.Ramgoolam for e-mail correspondence.
The research of T.T
was supported in part by JSPS Research Fellowships for Young
Scientists and also by DOE grant DE-FG02-91ER40654.

\appendix
\section{BPS Equation for D3-brane in PP-wave Background}
\setcounter{equation}{0}
\hspace{5mm}
Here we would like to derive the 
BPS equation for a D3-brane in the pp-wave background.
First we briefly review the supersymmetry which is
preserved by a Dp-brane
following the paper \cite{SkTa} (see also the references therein).
The projector $\Gamma$ ($\Gamma^2=1$) is defined as
\ba
(d\xi)^{p+1}\Gamma=-\f{1}{\s{-\det(G+F)}}\ e^{F}\we (\oplus_{n\in even}
\Gamma_{(n)}
K^nI),
\ea
where $\Gamma_{(n)}$ is given by
\ba
\Gamma_{(n)}
=(\d\xi^n\we\d\xi^{n-1}\we\ddd\we\d\xi^1)
\de_{i_1}X^{m_1}\de_{i_2}X^{m_2}\ddd\de_{i_n}X^{m_n}
\Gamma_{m_1 m_2 \ddd m_n}.\label{gn}
\ea
The operation $K$ is the complex conjugation and $I$ is
the multiplication
of $-i$. A spinor $\ep$ in the type IIB theory can be written as
$\ep=\ep_1+i\ep_2$, where $\ep_1$ and $\ep_2$ are the (16 components) 
Majorana-Weyl 10D spinor. The number of the preserved supersymmetries
is given by that of the Killing spinors $\ep$ which are invariant under
$\Gamma$.

Now we are interested in the maximally supersymmetric pp-wave 
(we set $\mu=1$ below for simplicity) 
\ba
ds^2=-2dx^+dx^- -(\sum_{a=1}^{8}(x_a)^2)(dx^+)^2
+\sum_{a=1}^{8}(dx_a)^2.
\ea
We denote the gamma matrix (as in (\ref{gn})) 
in this background as $\Gamma_{\mu}$ and one in the tangent 
space as $\gamma_{\mu}$.
Then we obtain 
\ba
&&\Gamma_{+}=\gamma_{+}+\f{|x|^2}{2}\gamma_-,\ \ \ \ \ 
\Gamma_{-}=\gamma_{-},\no
&&\{\gamma_+,\gamma_-\}=-2.
\ea
The Killing spinor $\ep$ in this background is given by
\ba
\ep&=&(1+\f{i}{2}\gamma_- (\sum_{a=1}^{4}x^a\gamma_a\gamma_{1234}+
\sum_{a=5}^{8}x^a\gamma_a\gamma_{5678}))\no
&&\times(\cos(x^+/2)-
i\sin(x^+/2)\gamma_{1234})(\cos(x^+/2)-
i\sin(x^+/2)\gamma_{5678})(\lambda+i\eta).
\ea
For convenience we also define 
\ba
\psi(x^+)=(\cos(x^+/2)-
i\sin(x^+/2)\gamma_{1234})(\cos(x^+/2)-
i\sin(x^+/2)\gamma_{5678})(\lambda+i\eta).
\ea

Let us turn to the main point. Consider the D3-brane whose world-volume
is $R\times R\times S^2$, where one of $R$s is the light-cone
time $x^+$
and another is the spatial direction $y(=x^4)$ and $S^2$ is 
the sphere\footnote{The relation to $R$ discussed in the context of 
D1-branes is given by $\ap R/2=\rho$.}$(x^1)^2+
(x^2)^2+(x^3)^2=\rho^2$.
We assume the radius 
$\rho$ depends only on $y$. In order to relate this system to 
$M$ non-abelian (expanded) D1-branes we assume the $M(\in Z)$ 
units of magnetic
flux on $S^2$ as $F=(\pi\al M)\sin\theta d\theta\we d\vp$,
where we introduce
the usual polar coordinate of the sphere. In this set up the projection 
is written as ($'$ means $\de_{y}$ and $\ap\equiv\pi\al M$)
\ba
\Gamma&=&\f{-1}{\s{(1+\rho^{'2})(y^2+\rho^2)(\rho^4+\ap^2/4)}}
(-\gamma_+ +\f{\rho^2+y^2}{2} \gamma_-)\no
&&\times(\gamma_{234}\rho^2+ \gamma_{123}\rho'\rho+\gamma_{4}K\ap/2
+\gamma_{1}\rho' K \ap/2)I.
\ea
Next it is easy to see that we must assume 
\ba
\gamma_{+}\psi(x^+)=0,\label{dys}\ \ \ \
(\mbox{or equally} \ \ \gamma_{+}\lambda=\gamma_{+}\eta=0),
\ea
which means that the preserved supersymmetries are all the dynamical
ones.
Then we obtain
\ba
\Gamma\ep&=&\f{i}{\s{(1+\rho^{'2})(y^2+\rho^2)(\rho^4+\ap^2/4)}}
(\f{\rho^2+y^2}{2} \gamma_- +i(\gamma_{234}\rho-\gamma_{123}y))\no
&&\times(\gamma_{234}\rho^2+\gamma_{123}\rho'\rho^2
-\gamma_{4}K\ap/2-\gamma_{1}\rho' K\ap/2)\psi.
\ea
Furthermore by using (\ref{dys}) we can find 
$(\gamma_{1234}+\gamma_{5678})\eta=(\gamma_{1234}
+\gamma_{5678})\lambda=0$.
Thus we can identify $\psi(x^+)$ with $\lambda+i\eta$.
Then the projection equation is rewritten as 
\ba
&&\f{1}{\s{(1+\rho^{'2})(y^2+\rho^2)(\rho^4+\ap^2/4)}}
(\gamma_{234}\rho-\gamma_{123}y)\no
&&\times (\gamma_{234}\rho^2+\gamma_{123}\rho'\rho^2
-\gamma_{4}K\ap/2-\gamma_{1}\rho' K\ap/2)(\lambda+i\eta)
=-(\lambda+i\eta).
\ea
or for each component
\ba
&&\Biggl[(y\rho'-\rho)\rho^2+\f{\ap}{2}(y+\rho\rho')\gamma_{1234}
+(y+\rho\rho')\rho^2\gamma_{14}+\f{\ap}{2}(y\rho'-\rho)\gamma_{23}
\Biggr]
\lambda, \no
&&=-\s{(1+\rho^{'2})(y^2+\rho^2)(\rho^4+\ap^2/4)}\lambda\no
&&\Biggl[(y\rho'-\rho)\rho^2-\f{\ap}{2}(y+\rho\rho')\gamma_{1234}
+(y+\rho\rho')\rho^2\gamma_{14}-\f{\ap}{2}(y\rho'-\rho)\gamma_{23}
\Biggr]
\eta \no
&&=-\s{(1+\rho^{'2})(y^2+\rho^2)(\rho^4+\ap^2/4)}\eta.
\ea

Finally we can conclude that the background preserves
eight (dynamical)
supersymmetries such that
\ba
\gamma_{1234}\lambda=\pm \lambda,\ \ 
\gamma_{1234}\eta=\mp \eta,     \label{1234}
\ea
if the following BPS equation is satisfied
\ba
2\rho^2(y+\rho\rho')=\pm \ap(y\rho'-\rho), \label{BPSEQ}
\ea
where we should choose\footnote{By using the BPS eq. we can
rewrite this
as $\ap y-2\rho^3<0$ ($\ap y+2\rho^3>0$).} $+\ (-)$ sign
if $y+\rho\rho'<0$ ($>0$). If we consider a D3-brane with
no-flux $M=0$ and
require the BPS equation $y+\rho\rho'=0$, 
then the background
preserves all of the dynamical supersymmetries (1/2BPS).
This solution is
given by the three dimensional sphere $y^2+\rho^2=const.$
and is no other than
the giant graviton as discussed in \cite{BeMaNa,SkTa,Pa}.

\end{document}